# The method of inhalation therapy with micro-doses of mixtures noble gases with oxygen.

A.Yu.Perov[1], B.M.Ovchinnikov[2]*, V.V.Parusov[2], A.V.Bobrovnikov[3], V.G.Safronova[4], Yu.D.Holodilin[5].

[1] - Hospital RAS, Moscow, doctor –anesthesiologist,

[2] - Institute of nuclear research RAS, Moscow, scientist,

[3] - Hospital Ekaterinburg, doctor –anesthesiologist,

[4] - Institute of cell biophysics RAS, Pushchino, scientist,

[5] - Plant "Tyumen aerosols", general director.

*Corresponding author; e-mail: ovchin@inr.ru

## Abstract

In this research, we presented the method of inhalation therapy with noble gas micro-doses and results of clinical studies in hospital of Russian Academy of Sciences (RAS). We have designed a device that allows dosed injection of noble gas with a volume of about 2-4 ml with peroxide vapors into the nasal cavity. After testing on the authors of the project and obtaining a positive impact of xenon and krypton, a study was conducted on a group of conditionally healthy volunteers in the amount of 20 people. We have noted a positive effect from 18 people. And also, in this article we briefly described the methods and devices that are used for inhalation therapy with mixtures of noble gases with oxygen.

## 1. Introduction

It has long been observed that all the noble gases (NG) that are used in industry, one way or another affect the person: xenon (xenon-oxygen mixture) has analgesic, antispasmodic, cardiotonic, neuroprotective, antistress, antihypoxic, immunostimulating, antiinflammatory, anabolic, neurohumoral, antispasmodic, regenerative, sedative effect [1].

Now there is a real boom in the use of xenon in anesthesiology [2], and, very carefully, in other specialties. Its use is very promising in cardiology, neurology, psychiatry, narcology, but its widespread use is constrained by a relatively high price. Although in our work [1] it is proved that the therapeutic effect is sufficiently low concentration, then the price becomes quite acceptable.

In our works [3-5] we have shown that krypton (krypton-oxygen mixture) has excellent healing properties. Moreover, these properties are superior to xenon in a number of indicators. If we consider that krypton is much cheaper than xenon, and the introduction of anesthesia at normal atmospheric pressure is not possible, it is relatively safe for wide use, followed by prospects for use in many sections of medicine.

Krypton has analgesic, antispasmodic, cardiotonic, neuroprotective, antistress, antihypoxic, immuno-stimulating, anti-inflammatory, anabolic, neurogomoral, myotonic, anticonvulsant, regenerative, tonic effect. These properties of krypton make it an effective tool for improving athletic performance of athletes, as it increases endurance and recovery of athletes after heavy physical activity, improves response. In clinical practice, krypton has shown itself as an effective tool for the treatment of post-stroke conditions.

Therapeutic use of the helium-oxygen mixture (heliox) was first reported in 1934 [6]. Density and viscosity of "heliox" are very different from those of air or oxygen. This can explain how "heliox" can induce modifications in the airway flow. In diseases of the main or small airways (upper airway obstruction, chronic obstructive pulmonary disease, asthma), such modifications could induce a diminution in the resistive component of the work of breathing and therefore protect against the risk of developing a respiratory failure. [7] The inclusion of inhalations of a heated gas mixture of helium with oxygen in the standard therapy of patients suffering from an infectious disease caused by SARS-CoV-2, with CT signs of pneumonia (CT2, CT3), with acute respiratory failure improves gas exchange, accelerates the elimination of the virus and indirectly increases the anti-inflammatory effect [8]. Therapy of the heated oxygen-helium mixture "heliox" (70% helium / 30% oxygen) is included in the recommendations in the complex intensive therapy of patients with COVID-19 in the initial stages of respiratory distress syndrome [9].

It has been experimentally proved that argon (argon-oxygen mixture) has the following properties: improves immune status, increases reproductive function, delays age physiological oppression of the sexual sphere, has antistress action*, is an analgesic and, most importantly, treats hypertension, which is very important, since 16.5% of all deaths (9.4 million per year) due to high blood pressure [10]. In Fig.1 we present of respiratory apparatus "DAMECA 10590" (left) and respiratory apparatus "Polynarcon-5" (right) for gas inhalation therapy.

*- depression, according to the world health organization, will become the second disease in the world after cardiovascular diseases by 2020, and there are no effective methods of its treatment [11]. Elimination of stress is necessary, because stress is the trigger of the onset of depression. Currently, the prevalence of depressive and anxiety disorders due to the impact of COVID-19 infection is repeatedly increased compared to that in previous years [12].

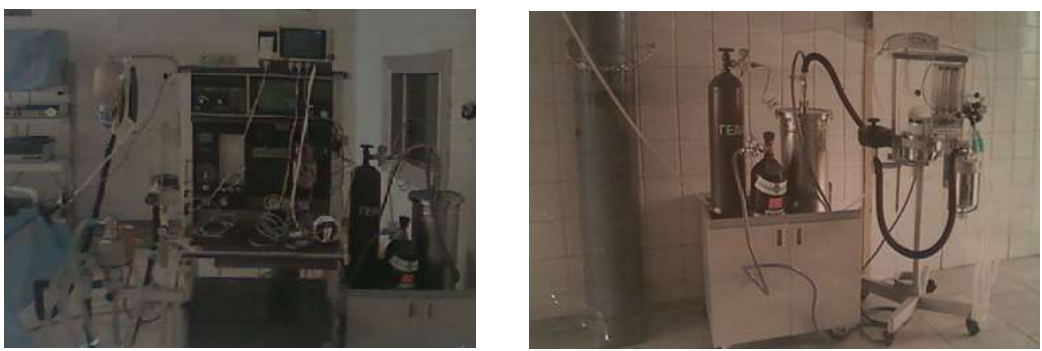

Fig.1. Respiratory apparatus "DAMECA 10590" (left), respiratory apparatus "Polynarcon-5" (right) in Hospital RAS with gas cylinders (argon, helium, xenon) and 20 liters absorber of xenon

## 2. Regenerative properties of noble gases

One of the main abilities (except for the effect of anesthesia), which is open in xenon, and also in argon, is the ability of these gases to activate the production of Hif-1 alpha protein in cells. Hif-1 alpha triggers the synthesis of many other biologically active proteins, including EPO (erythropoietin). And erythropoietin is the main protein that stimulates the regeneration of body tissues. This ability of noble gases explains many of their medicinal properties.

Noble gases cause the body to produce its own erythropoietin, and when used with argon, its own erythropoietin is in the hundreds, and even thousands cheaper than the injections. The effect is very powerful, it begins to work a day after inhalation, and lasts a long time. Below is a list of diseases for which the effective effect of inhalations of noble gases is revealed:

1. Hypertension. The effect of reduction (normalization) of arterial pressures.

2. Different types otolaryngic diseases. Otoprotective effect, improved blood supply to the nasopharynx. Effective treatment of respiratory diseases.

3. Improvement of capillary blood supply to the brain.

4. Improve and restore potency, improve the reproductive functions.

5. Improving the General condition of the body, the removal of stress.

6. Anti-inflammatory and immunoprotective effect.

7. Neuroprotective in patients more than 20 % increases cerebral blood flow, and renal, hepatic, etc. the bloodstream.

8. Radioprotective effect, xenon inhalations are used for recovery of vital organs after chemotherapy and radiotherapy.

9. The impact on the immune system, as a consequence, a decrease in the number of drugs used, including antibiotics.

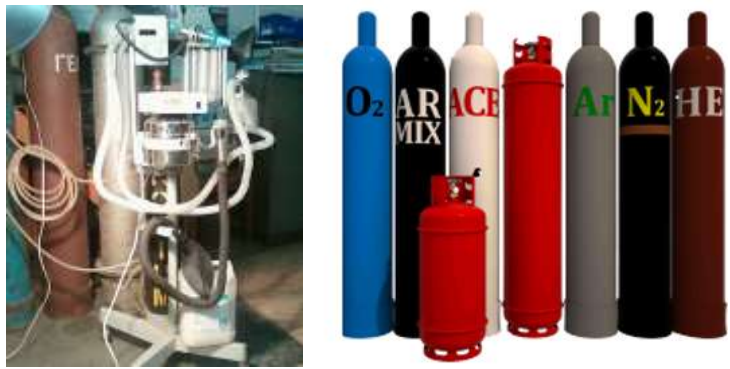

Fig.2. Respiratory apparatus "Polynarcon-5" with "INSOFT" gas-analyzer of oxygen in Institute for Nuclear Research of RAS (left) www.inr.ac.ru. Industrial gas cylinders (right)

And the list is far from complete, because the research continues and the list this one keeps expanding. But, unfortunately, all real clinical medical activity in Russia is only with xenon gas, because so far only he has one from the pharmacological Committee of the Russian Federation the permission to use.

There are studies on the effects of xenon inhalations on influenza virus with very encouraging results. There is a published and patented result for the treatment of hepatitis C by this technology using xenon gas. More than 30 people have been cured by this technology.

Regenerative properties of mixtures of noble gases with oxygen can be used for longevity of the elderly, because these mixtures restore the cells of the body. In Fig.2 shows "Polynarcon-5" respiratory apparatus with gas cylinders.

Currently, argon-oxygen mixtures were used in the hospital of the ministry of internal affairs of Yekaterinburg for the treatment of patients. As a result of therapy with these mixtures, wounds heal 3 times faster, without purulent deposits. In addition, the regenerative properties of these compounds can be used for longevity of the elderly, as these compounds improve the regeneration of body cells and have an antioxidant effect.

In Yekaterinburg opened a wellness center (Fig.3), which uses technology based on the use of noble gases for the treatment of cancer patients. These technology used when treating diseases such as oncological diseases, bacterial diseases, viral diseases, and benign tumors using therapeutic procedures conducted using breathing gas mixtures of oxygen and the noble gases argon, krypton, and xenon using an apparatus for creating a breathing gas mixture, and conducted inside of a device for blocking cellular memory in order to prevent pathogenic microorganisms from developing a tolerance for the treatment being conducted [13]. www.bobrovnikoff.com

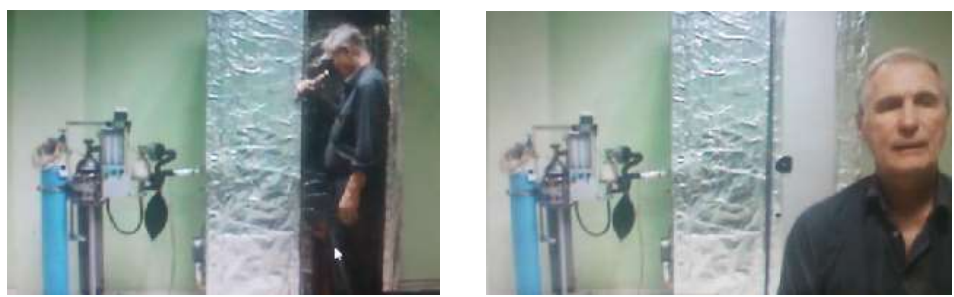

Fig.3. Respiratory apparatus "Polynarcon-5" with pressurized cabin in Yekaterinburg wellness center

## 3. Treatment of drug addiction and alcoholism

In our country, as well as in the US and Japan, drug addiction is treated by injecting a drug addict with xenon. In Yekaterinburg clinics drug addicts are treated with xenon - oxygen mixtures. For this purpose, we have successfully used therapy with both xenon-oxygen and argon-oxygen mixtures. In addition, we successfully used micro-doses of krypton-oxygen mixture.

## 4. Method of inhalation therapy with noble gas micro-doses (xenon, krypton, argon) with oxygen

When carrying out therapeutic inhalations on a closed circuit, the present medical staff and relatives noted that they are affected by micro-doses of xenon and krypton, up to the development of symptoms characteristic of therapeutic inhalation (increased efficiency, endurance, normalization of sleep, the appearance of liquid stool, etc.). Given that the concentration of noble gases in the ambient air is negligible, and with the appearance of leakage into the operating concentration increases slightly, but is accompanied by a significant impact on the conditionally healthy (including the authors), we have an assumption that for the manifestation of clinical signs enough micro-doses. The inhalation pathway through the lungs, apparently, is not necessary. The fact that noble gases have a powerful effect on the central nervous system and through it affect the entire body, has been proven earlier, including in our previous works for 2006-2008.

In order to increase and accelerate the permeability of xenon and krypton in the central nervous system, to enhance their biological effects, it was decided to use hydrogen peroxide as a transmitter [14], acting on the receptors of the vomeronasal organ – a miniature receptor organ located symmetrically in the nasal mucosa, which allows to weaken the "censorship" of the blood-brain barrier and provide penetration into the brain of a micro-dose of noble gas. It is known that the receptors of the vomeronasal organ are associated with the hypothalamus-an important center of coordination and control of vital functions of the body. In response to the information obtained, regulatory substances are synthesized in the structures of the hypothalamus — a kind of biochemical "reins" that stimulate or inhibit the production of hormones by the pituitary gland. The hypothalamus also synthesizes its own hormones that regulate water-salt metabolism, blood vessel tone, behavioral reactions, as well as the functions of the thyroid and gonads. Therefore, the effect of noble gas micro-doses seems to be manifested in the regulating effect on the hypothalamus.

We have designed a device that allows dosed injection of noble gas with a volume of about 2-4 ml with peroxide vapors into the nasal cavity. We used a simple version of device with pipetting of hydrogen peroxide (Fig.4).

After testing on the authors of the project and obtaining a positive impact of xenon and krypton, a study was conducted on a group of conditionally healthy volunteers in the amount of 20 people (mainly health workers of the Department of anesthesiology and intensive care). We have noted a positive effect from 18 people (increased performance, withdrawal symptom of chronic stress, thus reducing have a headache, have thus reducing symptoms of chronic allergic reactions, withdrawal syndrome has passed). The maximum multiplicity of administration was 2 times. Two volunteers received a negative result, which was due to the presence of chronic rhinitis and sinusitis in the subjects. After removing nasal congestion and cleansing of secretions, they also had an effect, but much less pronounced. The technique is in the stage of certification.

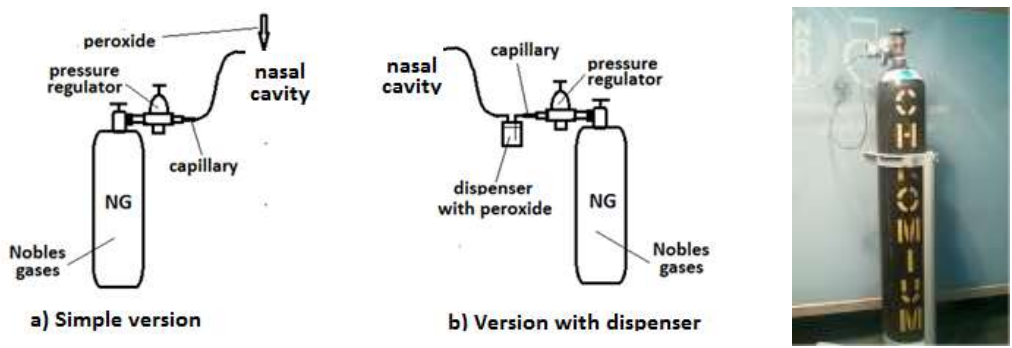

Fig.4 Two versions of device that allows dosed injection of noble gas with a volume of about 2-4 ml with peroxide vapors into the nasal cavity (left). Photo of krypton cylinder with simple version of device (right)

## 5. Industrial production of aerosol mixtures

Currently, the plant "Tyumen aerosols" established the industrial production of mixtures of argon with oxygen ($25\%Ar + 75\%O_2$) in cylinders made of aluminum with a volume of 1 liter with metered release of the mixture ($\sim 2 \div 3 см^3$) up to 150 times the volume of the mixture in the tank 17 liters [15]. Now begin to produce a mixture of krypton with oxygen ($25\%Kr + 75\%O_2$), which more effectively affects the body. As well as, the plant "Tyumen aerosols" established production of mixtures of $25\%Xe + 75\%O_2$. In Fig.5 shows gas aerosols of plant "Tyumen aerosols" and human gas exchange process. www.aerosol72.ru

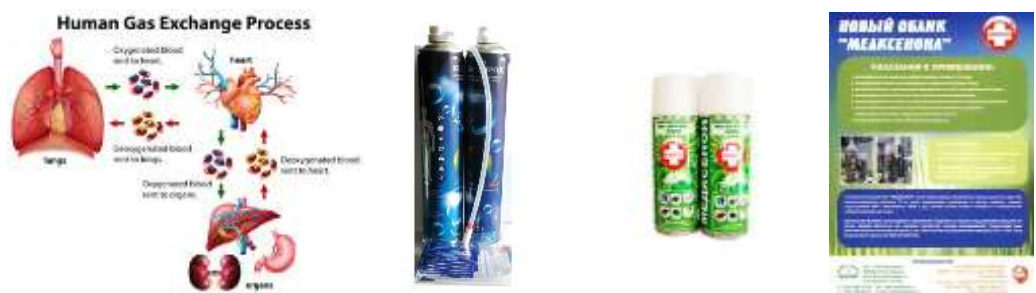

In Fig.5 The gas aerosols of plant "Tyumen aerosols" (right) and human gas exchange process (left).

## Conclusions

The fact that noble gases have a powerful effect on the central nervous system and through it affect the entire body, has been proven earlier, including in our previous works for 2006-2008 [1-5, 7]. In this study, we propose a method of inhalation therapy with micro-doses of mixtures noble gases with oxygen. In order to increase and accelerate the permeability of xenon and krypton in the central nervous system, to enhance their biological effects, it was decided to use hydrogen peroxide as a transmitter. In contrast to the treatment of mixtures of nobles gases with oxygen in a closed circuit using a stationary breathing apparatus, the method of treatment with micro-doses using cans with mixtures has the advantage that it is autonomous and can be used by any patient on their own, while it is safe and effectively complements the patient prescribed basic therapy.

The studied properties of helium suggest the effectiveness of helium-oxygen mixtures in patients with pneumonia caused by heat-sensitive coronavirus infection [8]. In this case, the state of hemodynamics is optimized, arterial hypoxemia is eliminated, microcirculation improves with an increase in the number of leukocytes and an increase in their phagocytic activity. This leads to dehydration, resorption of the inflammatory focus, more active delivery of various antibacterial drugs to the focus of infiltration [16]